# Analysis on the Mechanical Jamming of Particle Flow using Impeller-based Rheometer

Wenbin Xuan[1], Wenguang Nan[1*, 2]

1. School of Mechanical and Power Engineering, Nanjing Tech University, Nanjing 211816, China

2. Faculty of Engineering and Physical Sciences, University of Leeds, Leeds LS2 9JT, UK

*Contact Email: nanwg@njtech.edu.cn

**ABSTRACT:** We simulated the cohesive particle flow in an impeller-based rheometer using Discrete Element Method (DEM), and we focus on the dynamics of particles around the constriction between the blade and its surrounding vessel wall. The results show that mechanical jamming could transiently and intermittently occur in the constriction, but it is limited in a narrow region and short duration. Larger stiffness of particles and lifting flow pattern are more prone to the occurrence of jamming. The scaling law used to speed up the DEM simulation by reducing particle stiffness may fail for particle flow passing through clearance. The mechanical jamming of particles is in low frequency with value less than 50 Hz and the duration of an individual jamming event is usually less than 0.04 s. The existence of mechanical jamming is also illustrated by the experiment, where the wear of particle surface is clearly observed with scratches and pits.

# 1. Introduction

Dense granular flows are widely found in nature and industrial engineering, such as landslide, food processing, and additive manufacturing. The understanding and characterization of the rheological behaviour and dynamics of dense granular flow is highly desirable[1, 2], and it is now available by using commercial and in-house rheometers. Besides two co-axial vertical cylinders with powder sheared in the annular gap [3-5], a rotating blade has also been used in the rheometer of particulate system, including: 1) impeller-based powder rheometer of Freeman Technology [6, 7], where a twisted blade rotates while moving downward or upward through the particle bed; 2) powder cell of Anton Paar Modular Compact rheometer [8]; 3) in-house rheometer with a blade rotating at a specified height [9-11].  In the impeller-based powder rheometer of Freeman Technology, the granular flow is fully three-dimensional and transient, and it has the potential to transit from intermittent flow regime to quasi-static flow regime as the granular stress varies during the process of the blade penetrating into the bed, as reported by Nan et al. [12, 13]. This is much different from the Couette device  [3-5] and the latter two kinds of rheometer which work under constant normal stress, making the impeller-based powder rheometer of Freeman Technology being attractive for the analysis of the rheological behaviour of granular flow. Till now, the effects of particle properties (cohesion and shape), fluid medium and strain rate on the rheology behaviour of granular flow in this rheometer have been explored by various researchers [7, 12, 14, 15] through the analysis of the stress field of bulk particles. Recently, Nan et al. [13] identified the existence of non-local effects of blade shearing in this rheometer, and they proposed a mathematical model to unify the transient viscosity by introducing the concept of granular temperature in kinetic theory. However, as the rotating blade with a twisted shape moves through the particle bed, the small gap between the blade tip and surrounding vessel wall may induce jamming if the particle size is comparable to the gap size. This has not yet been reported and analysed.

Meanwhile, reliable control and prediction of granular flow through constrictions is of great interest in particulate processing industries. However, a major problem is particle jamming, which makes the granular flow become non-continuous and even halted. A common example is the particle jamming in the particle feeding and conveying systems, including hopper flow driven by gravity [16-18], belt flow driven by friction [19, 20], pipe flow driven by fluid [21, 22], where the jamming is usually in static equilibrium with granular flow permanently halted. Recently, another kind of jamming is reported by the numerical work of Nan et al. [23] using Discrete Element Method and their following experimental work [24] for blade spreading system, where a particle heap is spread onto a rough baseplate by a blade through a gap (i.e. the clearance between the moving blade and stationary baseplate) with size comparable to particle diameter, resulting in particles being transiently and intermittently jammed under the combined effect of blade shearing and stationary wall. Besides the gap size, the jamming in the latter one is affected by the shear action of the blade, and it involves the sudden change of the passage of granular flow, making the prediction of the rheological behaviour of particles much more difficult than the first one. Till now, little work has been made on the physics of transient and intermittent jamming of granular flow through constrictions. As a commercial rheometer, the impeller-based powder rheometer of Freeman Technology has been widely used in industry and academia nowadays for assessing powder flowability and predicting particle rheology in a comparative way. Although this rheometer also has the potential to investigate the jamming of particles through constrictions, it has not been reported so far.

In this work, the mechanical jamming of granular flow through constriction is analysed by simulating the dynamic behaviour of cohesive particles with size comparable to the constriction in an impeller-based powder rheometer with discrete element method (DEM). By analysing the effects of particle stiffness and particle cohesion on the transient force and torque of the blade and bulk flowability, as well as the characteristics of the jamming structure and jamming events, the physics of mechanical jamming is explored, following which the effect of mechanical jamming on the particle surface damage is experimentally discussed. The results will have a notable impact on the further understanding of the nature of jamming in particulate system, and the rheology of granular flow through constrictions.

## 2. Methods

In this work, the granular flow in the impeller-based powder rheometer of Freeman Technology is simulated by Discrete Element Method (DEM), where the particles are modelled as discrete phases and their motions are tracked individually by solving Newton's laws of motion [25, 26], for which Altair EDEM™ software package is used. The simulated system comprises a cylindrical glass vessel with diameter of 25 mm, and a blade moving up or down while rotating inside a particle bed contained in the vessel [12], as shown in Fig. 1. The blade with a twisted shape is manufactured from hardened stainless steel, and its diameter is 23.5 mm, resulting in a minimum gap of 0.75 mm between the vessel wall and the blade tip, as shown in Fig. 1(e). The vertical translational speed $u$ and rotational speed $\omega$ of the blade are given as:

$$u = u_{tip}\sin(\alpha) \tag{1}$$

$$\omega = u_{tip}\cos(\alpha) / R_{blade} \tag{2}$$

where $\alpha$ is the helix angle; $u_{tip}$ is the tip speed; $R_{blade}$ is the blade radius. The elastic contact force between particle and particle is described by Hertz-Mindlin contact model [26], and the adhesive interaction is accounted for by JKR theory [27], in which the normal contact force is given as:

$$F_n = \frac{4E^* a^3}{3R^*} - \sqrt{8\pi \Gamma E^* a^3} \tag{3}$$

where $\Gamma$ is the interfacial surface energy; $E^*$ is the equivalent Young's modulus; $R^*$ is the equivalent radius of particle; $a$ is the contact radius, which can be calculated from the normal overlap $\alpha$:

$$\alpha = \frac{a^2}{R^*} - \sqrt{\frac{2\pi \Gamma a}{E^*}} \tag{4}$$

In the unloading process, the normal contact force $F_n$ is not zero when the normal overlap $\alpha$ is negative, as further work is required to separate the cohesive contact. For simplicity, the damping force and tangential contact force are not shown here, which can be referred to Thornton [26].

Spherical glass beads with mean diameter of $d=0.925$ mm are used in this work, and the bed is randomly generated using a uniform size distribution of particles in the range of $0.8d$-$1.2d$. Prior to the moving of the blade, particles above the pre-set height of 52 mm are removed, resulting in a uniform particle bed with a smooth surface. The tip speed $u_{tip}$=0.25 m/s is used in this work, and two successive tests are carried out: 1) the blade rotates anti-clockwise while moving down from the bed free-surface to the bed bottom, i.e. $\omega$=21.2 rad/s and $u$=−21.8 mm/s, resulting in high stress and compression within the granular bed; 2) the blade rotates clockwise while moving up from the bed bottom to the bed free-surface, i.e. $\omega$= −21.2 rad/s and $u$=21.8 mm/s, resulting in low stress and lifting flow pattern within the granular bed. The material properties and interaction parameters of glass beads are summarized in in Table 1 and Table 2 based on the experimental characterization done by Pasha et al. [28] for single glass bead. In the standard case (case A), the interfacial surface energy of glass bead is 0.274 J/m$^2$. To explore the effects of the stiffness and cohesion of particle on the dynamics of granular flow and jamming of particles in the rheometer, another two cases are considered here, as summarized in Table 3: 1) case B, Young's modulus is reduced to 63 MPa, while the interfacial surface energy is unchanged, resulting in the same Bond number as the standard case; 2) case C, Young's modulus is reduced to the same value as case B, but the interfacial surface energy is correspondingly scaled down to 0.017 J/m$^2$ to obtain the same cohesion number as case A. Here, the Bond number is the ratio of the maximum tensile force predicted by JKR theory to the particle weight ($mg$), and the cohesion number is the ratio of the adhesive work to the particle's gravitational potential energy with a characteristic height equal to particle radius, given as [29, 30]:

$$Bo = \frac{1.5\pi\Gamma R}{mg} = \frac{9}{8}\frac{\Gamma}{\rho R^2 g} \tag{5}$$

$$Coh = \frac{1}{\rho g}(\frac{\Gamma^5}{E^2 R^8})^{1/3} \tag{6}$$

where $E$ is the Young's modulus, $\rho$ is the density, and $R$ is the particle radius. For all three cases, both downward test and following upward test are conducted. According to the scaling law, bulk particles in case A should have the same flowability as case B or case C. The timestep in simulation is kept the same in all cases, i.e. 5×10$^{-8}$ s, which is around 12% of Rayleigh timestep, and the data is exported at the frequency of 500 Hz.

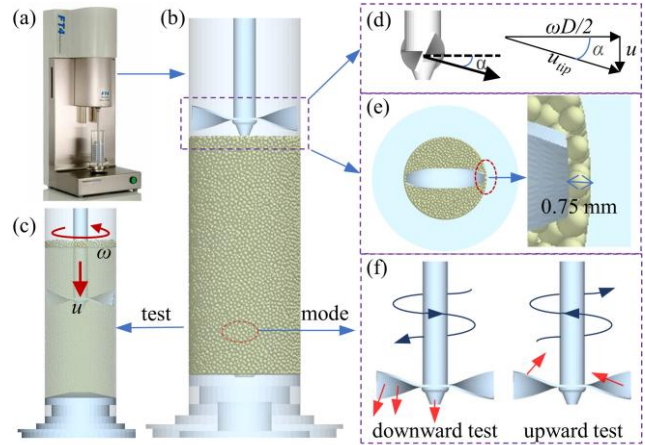

Fig. 1. Schematic illustration of (a) impeller-based powder rheometer of Freeman Technology, (b) simulation system, (c) downward motion with anti-clockwise rotation of the blade in the downward test, (d) helix angle of velocity vector in the downward test, (e) gap between the blade tip and cylindrical vessel wall, and (f) motion of the blade in downward test and upward test.

Table 1. Material properties of glass beads and geometry.

| Material property | Particle/Vessel | Blade |
|---|---|---|
| Density, $\rho$ (kg/m$^3$) | 2500 | 7800 |
| Young's modulus, $E$ (GPa) | 63 | 210 |
| Poisson ratio, $v$ | 0.2 | 0.3 |

Table 2. Contact interaction parameters of glass beads and geometry.

| Interaction property | Particle-particle/vessel | Particle-blade |
|---|---|---|
| Restitution coefficient, $\mu$ | 0.9 | 0.9 |
| Friction coefficient, $e$ | 0.5 | 0.3 |
| Rolling friction, $\mu_r$ | 0.01 | 0.01 |

Table 3. Parameters in cases A, B and C.

| Parameters | Case A | Case B | Case C |
|---|---|---|---|
| Young's modulus, $E$ (GPa) | 63 | 0.063 | 0.063 |
| Surface energy, $\Gamma$ (J/m$^2$) | 0.274 | 0.274 | 0.017 |
| Bond number | 58.8 | 58.8 | 3.7 |
| Cohesion number | $2.3\times10^{-4}$ | $2.3\times10^{-2}$ | $2.3\times10^{-4}$ |

## 3. Results

### 3.1. Flow resistance

Fig. 2 shows the evolution of the flow pattern of particles with time, where the particles are coloured based on their velocity magnitude. As the rotating blade penetrates into the particle bed, the particles at the free surface of the packed bed are firstly agitated, resulting in a cascading pattern of particle flow around the blade. This is followed by a compressive and shear flow of particles imposed by the blade, where particles with large velocity are mainly limited in front of the blade, whist particles away from the blade are almost stationary. There is almost no distinct difference between case A and case B in terms of the flow pattern of particles, suggesting that the granular flow in the rheometer considered in this work is mainly affected by the blade motion.

Fig. 3 shows the variation of the total force and torque imposed on the blade in the vertical direction as the rotating blade impenetrates into the particle bed. In case A, the force could jump to above 40 N while the torque could suddenly increase to around 2 N·m. They are much larger than the data reported in previous work [28, 31]. Thus, mechanical jamming is expected to occur, in which the clearance (i.e. the gap between the blade tip and the cylinder's inner wall in this work) is locally blocked by a few of particles, and an indicator is the extremely large force arose from the interaction between particles and geometry walls, as strong contact force network would be formed to bear the shear action of the moving blade or the stagnation effect of the stationary vessel wall. However, this kind of jamming is not stable as the blade continues to move. The strong contact force network would be broken and disappear under the shear and compressive action of the blade, and the corresponding force and torque of the blade quickly decrease to the value before the occurrence of jamming. Therefore, the jamming is transient, and it could occur and disappear frequently, resulting in a number of pulse-style peaks of force/torque in case A, as observed in Fig. 2(a) and (c). It seems that there is no clear trend for the value and time location of the peaks, indicating that the jamming is not sensitive to the penetration depth of the blade within the bed.

Compared to case A, the force and torque of the blade in cases B & C are much smaller. For example, the maximum value of the force and torque in case B is only 1/20 and 1/100 of that in case A, respectively. Meanwhile, the force and torque of the blade in cases B & C increase monotonously with time as the rotating blade moves downward, although there exist some small fluctuations. It suggests that almost no significant mechanical jamming occurs in cases B & C. Of course, the fluctuations of the force and torque of the blade in cases B & C may also indicate the occurrence of jamming events, but those potential events would be much weaker than that of case A, thus, they are not focused here. Thus, the mechanical jamming is very sensitive to the particle stiffness, and larger stiffness is more prone to the occurrence of jamming. Meanwhile, the variation of the force/torque with time in case B is almost similar to that of case C, indicating that particle cohesion has a smaller effect on the resistance of particle flow to blade action than that of mechanical jamming. It is interesting that although jamming occurs in case A, its flow pattern of particles does not show much difference from that of case C. This suggests that the jamming is limited in a very narrow region, and the duration of each jamming event is also very low, resulting in almost no significant effects on the whole structure of granular flow. It should be noted that the peaks hide the overall trend of the variation of force and torque with time in case A. If re-plotting the absolute value of the force/torque in the log scale, the overall trend of the curves in case A would be more clearly observed, which is similar to that of cases B & C, as shown in Appendix. It demonstrates again that the effect of jamming on granular flow is limited in a narrow region and short duration, and the jamming almost has no influence on the dynamics of granular flow in following time steps.

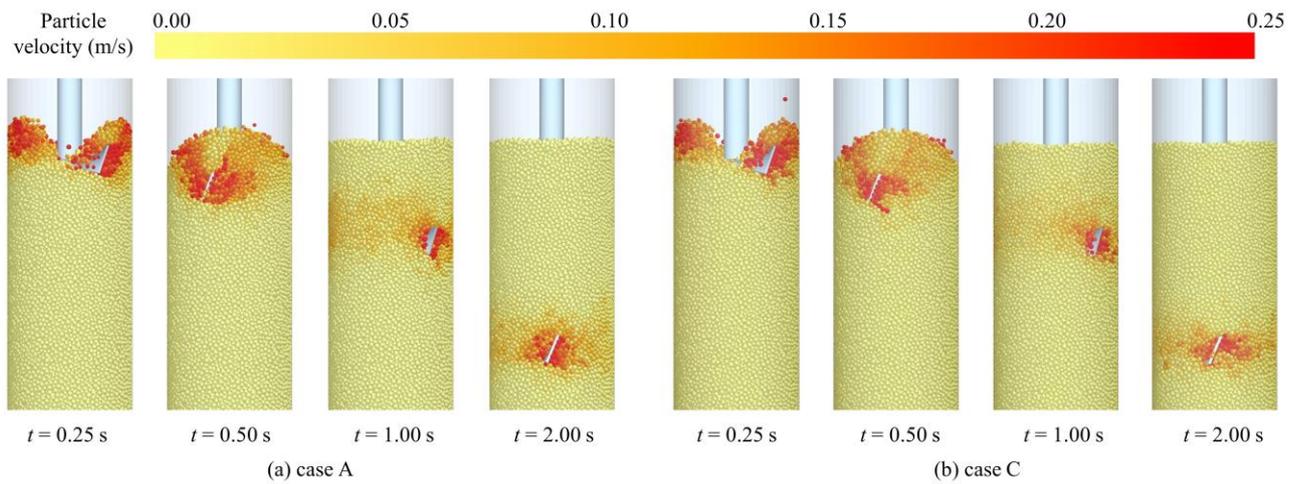

Fig. 2. Flow pattern of particles in the downward test in the rheometer for case A and case C.

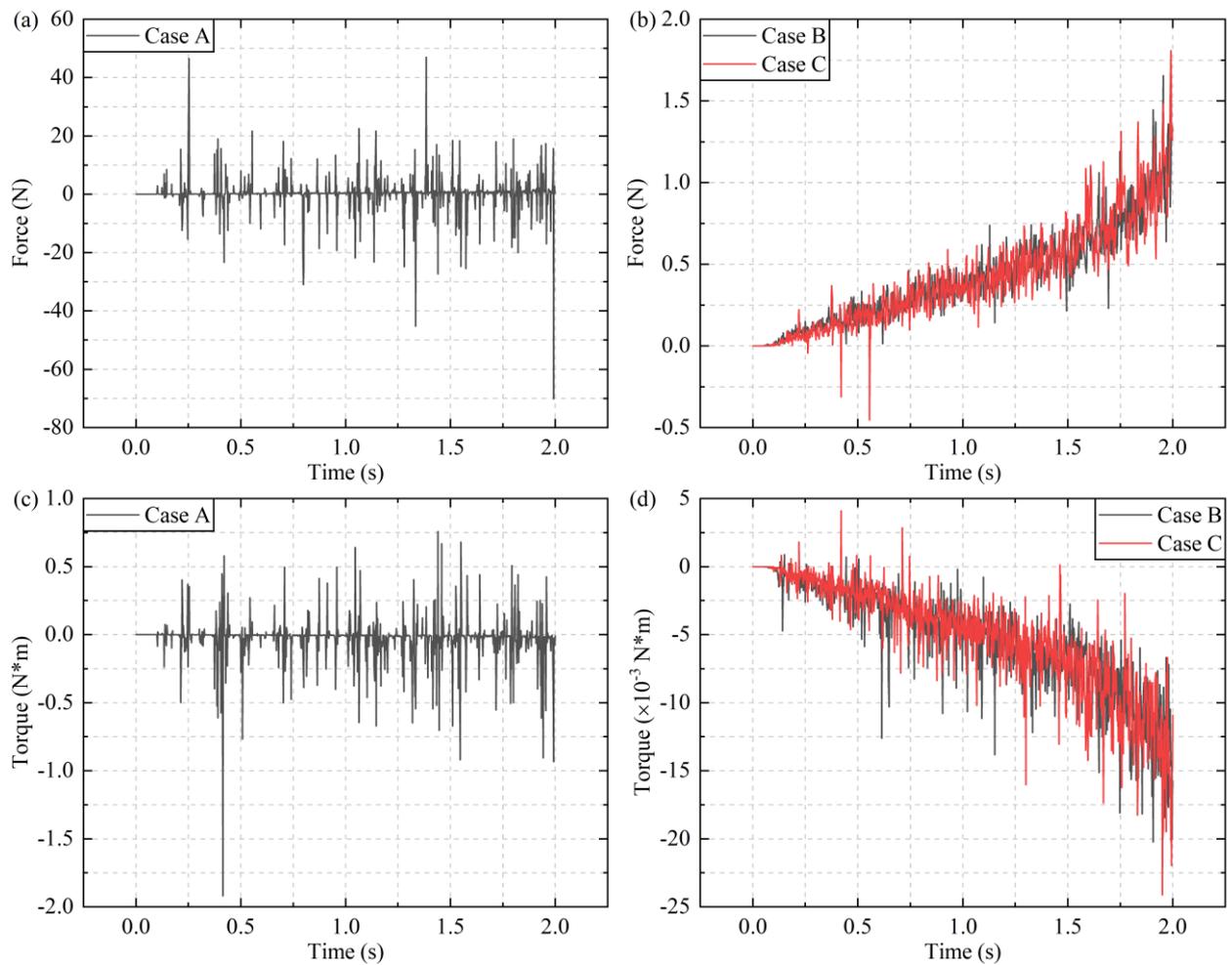

Fig. 3. Variations of the total force and torque of the impeller in vertical direction with time: (a) case A and (b) cases B and C for force, (c) case A and (d) cases B and C for torque, where only the results in the downward test are shown for simplification.

The resistance of particle flow to the blade motion, i.e. the mechanical work done by the blade, is referred to as the flow energy. Based on the axial force $F$ and the torque $T$ of the blade as illustrated above, flow energy *FE* is given as:

$$FE = \int_0^H \left( \frac{T}{R_{blade} \tan \alpha} + F \right) dH \qquad (7)$$

where $H = u \cdot t$ is the penetration depth. The larger the flow energy, the more difficult the particles flowing. Fig. 4 shows the variation of flow energy with time in both downward and upward tests. In cases B & C, the flow energy in the downward test increases monotonically with time in a power law, which is consistent with previous work by Nan et al. [12]. However, as shown in Fig. 4(a), the variation of flow energy with time in case A deviates significantly from this behaviour. In case A, the flow energy increases with time in a stepwise/climbing style, i.e. there are large and sharp increments of flow energy at a number of time points. This is due to the mechanical jamming discussed above, in which the force and torque imposed on the blade jump suddenly, as shown in Fig. 3(a) and Fig. 3(c). Meanwhile, the jamming induced flow energy is much larger than that of the cases without jamming. For example, the maximum flow energy in the downward test of case A is about 1.6 J, which is about 10 times larger than that of cases B and C. Therefore, if jamming occurs, the flow energy could not represent the flowability of bulk particles anymore, as the flow energy is significantly disturbed by the jamming instead of the intrinsic cohesive and frictional nature of bulk particles. Meanwhile, the results also demonstrate that the scaling law [32] used to speed up the simulation of particle flow fails if the jamming potentially occurs, in which the bond number or cohesion number is kept constant when decreasing the particle stiffness to gain a larger timestep in DEM simulation.

As shown in Fig. 4(b), in cases B & C, the total flow energy (i.e. $t=2.0$ s) in the downward test is larger than that of the upward test. This is intuitively expected as the granular flow is in a compressive mode in the downward test while it is in a lifting mode in the upward test. However, in case A, as shown in Fig. 4(a), it is interesting that the transient flow energy in upward test is larger than that of downward test at most time. It demonstrates that the jamming in the upward test is stronger than that of downward test. It also suggests that if mechanical jamming repeatedly occurs during the test, the flow energy is more affected by the mechanical jamming instead of the test mode or the bulk stress within the particle flow. Meanwhile, the flow energy in case B is larger than that in case C in the downward test, which is intuitively expected as the particles in case B have a larger interfacial surface energy, resulting in larger cohesion within bulk particles. The difference of flow energy between case B and case C is enlarged in the upward test, indicating that the dynamic bulk cohesion could be better characterized in a lifting and soft flow pattern.

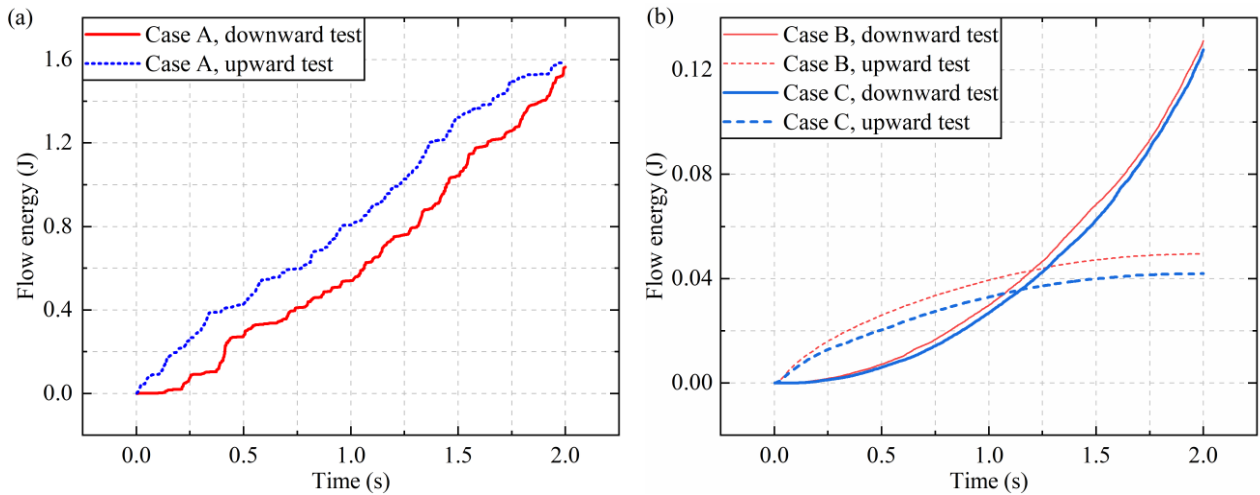

Fig. 4. Variation of flow energy with time, (a): case A, (b): cases B & C.

**3.2. Jamming structure**

Based on the jamming observed from the existence of extremely large peaks of blade force and torque shown in Fig. 3, the particles which are jammed could be further examined through the analysis of compressive force of particles. Fig. 5 illustrates the variation of compressive force of particles with time for case A, where both the averaged and maximum value are presented, and case B is included for comparison. The results show that the maximum value of compressive force is much larger than its averaged value, indicating a significantly non-uniform distribution of inter-particle force, especially in case A. This is expected as the granular flow is mainly locally sheared around the blade, as shown in Fig. 2. Meanwhile, the maximum compressive force in case A is in the magnitude of kN, which is much larger than that in case B with the magnitude of N. This is caused by the mechanical jamming as reported in Fig. 3. It should be noted that under this large compressive force, the particles are actually under plastic deformation and surface damage, resulting in the decrease of the resistance to blade shear, which is not considered in this work. Here, by comparing the maximum compressive force of particles between case A and cases B & C, a critical value of 20 N is selected as the criteria for the occurrence of mechanical jamming, i.e. the particles with compressive force larger than this critical value are deemed as jammed particles. In this way, the detection of jamming events using particle compressive force would agree well with the ones using the blade force and blade torque shown in Fig. 3, in which significant mechanical jamming is only observed in case A.

Fig. 6 shows the snapshots of the characteristic particles which are jammed, while un-jammed particles are hidden and not shown. The results show that jammed particles are almost all located in the gap between the blade tip and vessel wall. However, the exact locations of jammed particles are almost random, and the jamming can occur at any positions along the blade tip edge. For example, the jammed particles occur in the corner of the blade edge, as shown in Fig. 6(d), and they can also occur in the middle of the blade edge, as shown in Fig. 6(e). Meanwhile, the number of jammed particles is also not fixed. For example, there are 4 jammed particles in Fig. 6(b) while only 1 jammed particle in Fig. 6(e). However, based on the statics of the number of jammed particles at different time, the jamming structure consisting of only 1 jammed particle has the largest probability, which is shown in Appendix.

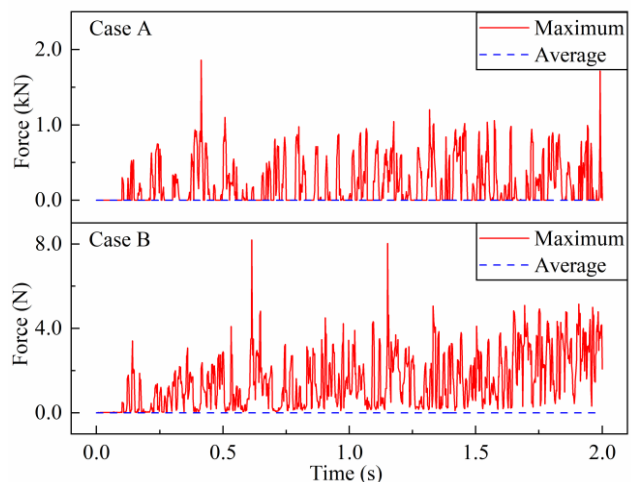

Fig. 5. Variation of the maximum and average value of particle compressive force with time, where only the downward tests are included for simplification.

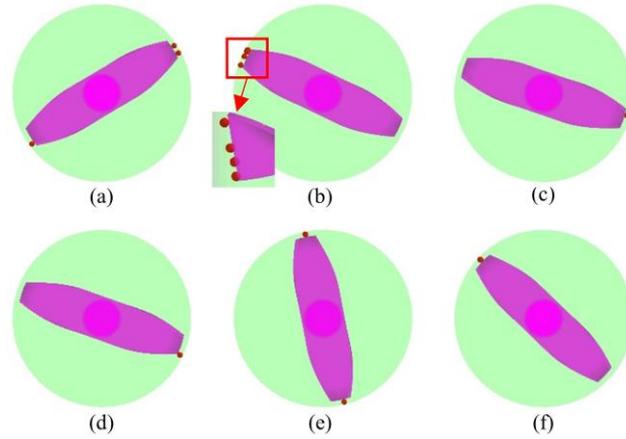

Fig. 6. Snapshots of the jammed particles in case A: I) downward test: (a) 0.618 s, (b) 1.016 s and (c) 1.318 s; II) upward test: (d) 0.238 s, (e) 1.176 s and (f) 1.740 s.

To illustrate the evolution of the jamming structure, an individual jammed particle is tracked. Its spatial position and velocity, as well as the snapshots of contact force network with jammed particle coloured by purple, are shown in Fig. 7. The results show that the jammed particle is almost squeezed into and out of the gap, with a trajectory across the blade edge. The contact force network could only resist the load from blade action in the rotational direction. Meanwhile, strong contact force network is formed within the gap, connecting the jammed particle to the blade edge and the vessel wall. However, the strong contact force network only survives for a short time and disappears quickly as long as the particle is not jammed anymore. For example, as shown in Fig. 7(a), the strong contact force network forms between 1.314 s and 1.318 s, while it disappears between 1.322 s and 1.326 s. Based on the velocity distribution of jammed particles and surrounding particles, it could be found that the jammed particles could be affected by the blade, with the velocity comparable to blade speed (i.e. 1.318 s in Fig. 7(a)), or stagnated by the vessel wall, with almost zero velocity (i.e. 1.740 s in Fig. 7(b)). It demonstrates that both the stagnation effect of stationary vessel wall and the shear effect of the moving blade have effects on the jamming structure. It is predicted that the roughness of vessel wall would affect the formation and disappearance of jamming event. This kind of jamming is different to the intermittent jamming described by stress-strain relation in the quasistatic shearing system of bulk frictionless particles, as simulated by Heussinger [33], where the jamming is mainly controlled by the critical particle volume fraction.

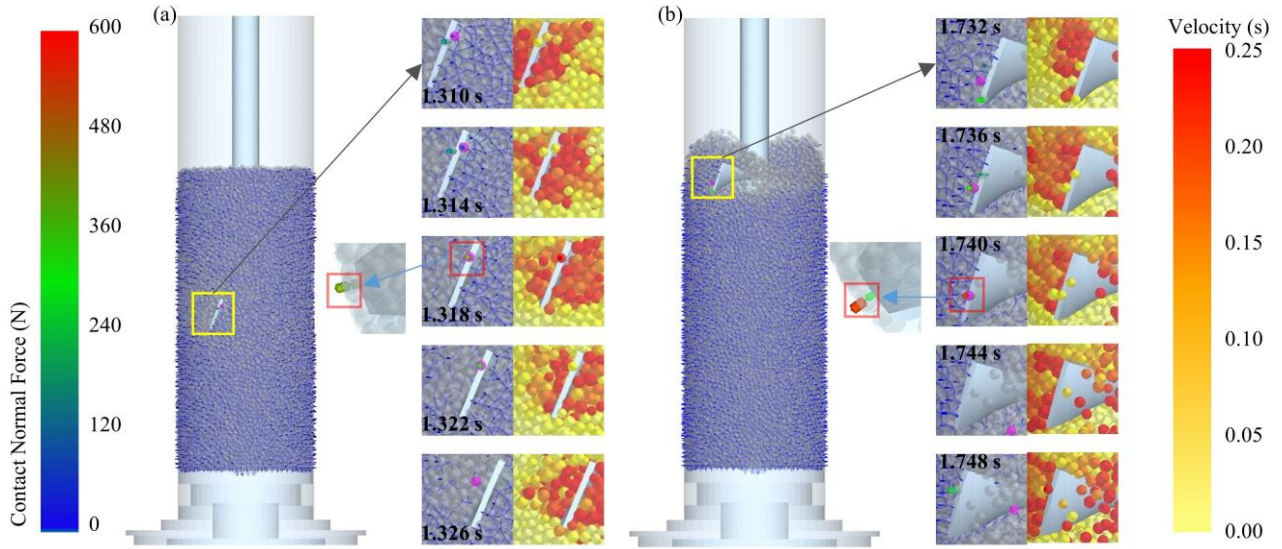

Fig. 7. Snapshot of evolution of the jamming event in (a) downward test and (b) upward test, where both contact force network (jammed particle is coloured by purple) and velocity of particles are included.

Based on the evolution of maximum compressive force of particles with time, the jamming event could be detected one by one based on the algorithm schematically shown in Fig. 8. Continuous data points with compressive force larger than critical value (20 N) are deemed to belong to the same jamming event. A big jamming event can include several force peaks, as clearly observed from the shadow region in Fig. 8. Here, the characteristics of jamming events are quantified by the survival time and frequency. The survival time is defined as the duration between the start and end of the jamming event, as illustrated by the red dots in Fig. 8. The corresponding frequency is calculated by dividing the number of jamming events by the total simulation time (i.e. 2 s). The jamming events are classified into 6 classes based on their survival time, and the total frequency of each class with the corresponding survival time is shown in Fig. 9. The results show that the frequency decreases sharply with the survival time, i.e. the longer the jamming event, the less frequency of it. Meanwhile, except for the jamming events with $\Delta t<10^{-2}$ s and $\Delta t=(2\sim3)\times10^{-2}$ s, the frequency of others in the upward test is larger than that in the downward test. Meanwhile, the total duration of jamming events in the upward test is longer than that in the downward test. Thus, the jamming events may be easier to occur in the upward test. It maybe due to that in the upward test, the granular flow is in soft and lifting flow pattern, and therefore, the jamming structure could resist more to the blade motion to some extent. Fig. 9 also shows that the total frequency of jamming events in the rheometer detected from the force is in low frequency, i.e. less than 50 Hz. To illustrate the contribution of each class of jamming events to the total duration of jamming, the summation of the survival time of jamming events belonging to this class is calculated and divided by the total duration of all classes of jamming events, as shown in Table 4. The largest probability is observed for the jamming event with a survival time of $\Delta t=(1\sim2)\times10^{-2}$ s for both tests, while the 2nd largest probability is found at $\Delta t=(2\sim3)\times10^{-2}$ s and $\Delta t=(3\sim4)\times10^{-2}$ s for the downward and upward tests, respectively. Table 4 also shows that single jamming event with duration less than 0.04 s are dominant with a total probability of around 90%.

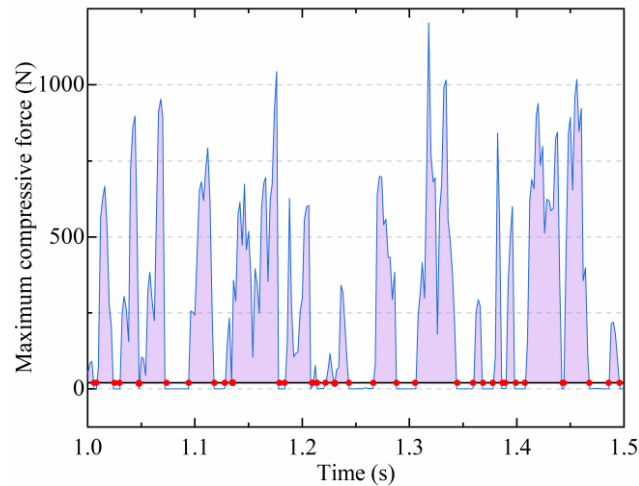

Fig. 8. Example of the detection of the jamming event.

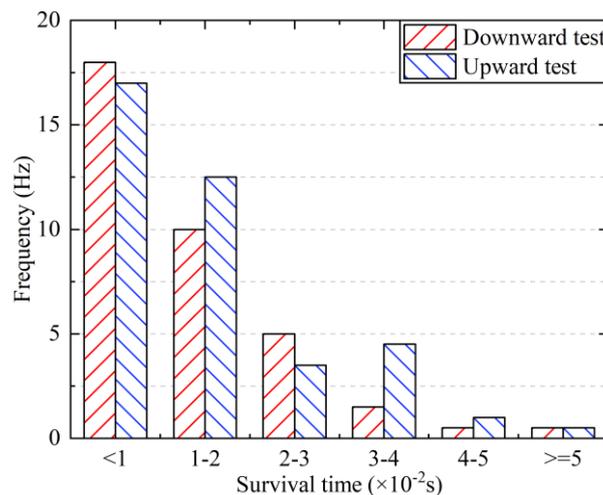

Fig. 9. Variation of the frequency with the survival time for the jamming events in case A.

Table 4. Probability of jamming events at the corresponding duration.

| Survival time (s) | Probability (%) | |
| --- | --- | --- |
| | Downward test | Upward test |
| <0.01 | 19.0 | 14.7 |
| 0.01-0.02 | 32.8 | 30.2 |
| 0.02-0.03 | 25.1 | 13.5 |
| 0.03-0.04 | 11.5 | 28.8 |
| 0.04-0.05 | 5.4 | 8.2 |
| >=0.05 | 6.3 | 4.6 |

## 4. Discussions

Fig. 10(a) shows the SEM image of spherical glass beads after a number of standard downward tests in the FT4 powder rheometer of Freeman Technology. The number based $D_{90}$ of glass beads is 360 μm as characterised by Mastersizer 2000, resulting in a normalised gap size of $\delta/D_{90}=2.1$. Several scratches and pits could be observed on the surface of glass beads. This is clear evidence of mechanical jamming from experiment. Under the effect of jamming, the particle is subjected to large force from the blade edge/tip, as analysed from simulations shown above, and a small quantity of material is even dug out, resulting in several pits with different depth and size on the surface of glass beads. However, as shown in Appendix, for the total force and torque of the blade in the vertical direction, the extremely large peaks like the ones in DEM simulation for large particles in Fig. 3, are not observed in the experiment. It may be due to that the data of measured force and torque is auto filtered and smoothed by the control system of the rheometer to reduce the fluctuation, then it is exported at a frequency of 25 Hz (i.e. 0.04 s) as used here. Under this condition, potentially large peaks like the ones in Fig. 3, which are reported to survive for a very small interval (less than 0.04 s), would be hidden. Therefore, to detect the occurrence of jamming in terms of the force and torque imposed on the blade in the experiment, the sensors with higher data acquisition frequency may need to be installed within the rheometer.

It should be noted that when using large glass beads in the experiment, such as particles with number based $D_{90}$ larger than 600 μm, very large mechanical noise could be intermittently heard, making the rheometer at the risk of damage and the experiment could not be continued. The maximum force and torque among all test methods of FT4 rheometer instrument, including shear cell test and compressibility test, are reported to be 50 N and 0.9 N·m in the user manual [34], respectively. For the standard downward tests reported here, the allowable values would be smaller as the blade is involved. Meanwhile, the jammed particles are under plastic deformation or surface damage, which have not yet been considered in the DEM simulations. They are the reasons why the experiment could not be compared directly to the simulations. However, both the findings in the simulation and experiment validate the existence of mechanical jamming of granular flow through the gap between the moving blade and stationary vessel wall in the rheometer when the size of the particles is comparable to that of the clearance. The sensitivity of jamming to particle stiffness is also observed in the experiment, as shown in Fig. 10(b), in which the polyethylene spheres with number based $D_{90}$ = 628 μm are used and they have lower stiffness than glass beads. The SEM images demonstrate again that the mechanical jamming would cause damage and wear on particle surface, in which several long and fine scratches as well as large patches are significantly observed on the particle surface. The extent of surface damage is weaker than that of glass beads due to weaker jamming under the effects of lower Young's modulus. However, it should be noted that the situation of mechanical jamming is only for the powder sample which has some stiff particles with size comparable to the gap between impeller tip and cylinder's inner wall, i.e. number based $D_{90}$ of powder sample is comparable to gap size. Of course, the jamming in the rheometer could be avoided by using the cylinder vessel with larger size or the blade with smaller diameter, in which the gap size between the blade tip and cylinder's inner wall is enlarged. Based on the findings from DEM simulations and experiments, it could be inferred that mechanical jamming would be minimised to some extent when the gap is larger than $2.5D_{90}$ for the powder sample with small particle size or much less than $D_{10}$ for the powder sample with large particle size, where $D_{10}$ and $D_{90}$ are number based diameters of powder sample. The jamming would also be weakened if the particles with low stiffness are adopted.

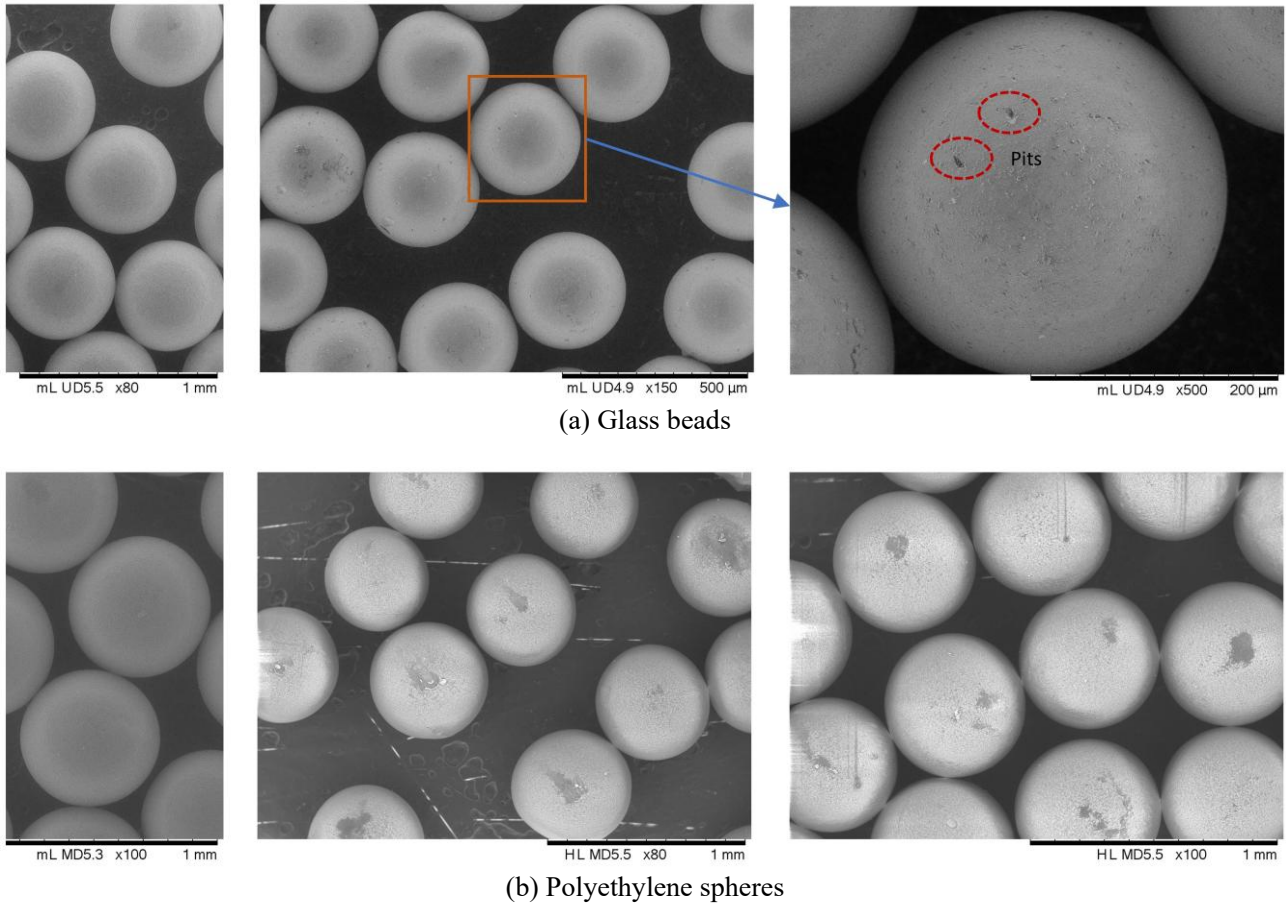

(a) Glass beads

(b) Polyethylene spheres

Fig. 10. SEM images of the surface damage and wear of particles caused by mechanical jamming: the left images are before the test; the middle and right images are after the test.

## 5. Conclusions

The mechanical jamming of granular flow in an impeller-based powder rheometer is studied using DEM simulation. The force and torque on the blade, as well as the flow energy used to characterize the flowability of bulk particles, are analysed. This is followed by the investigation of the jamming structure and the characteristics of jamming events. The existence of mechanical jamming is also validated by the experiment. The main results of this work are summarized as follows:

1) The mechanical jamming in the impeller-based powder rheometer is transient, and it could occur and disappear frequently. The jamming is limited in a narrow region and short duration. Jamming is more serious in the test mode with soft and lifting flow pattern, and it is sensitive to particle stiffness, where larger stiffness is more prone to the occurrence of jamming.

2) The flow energy used in the rheometer could not represent the flowability of bulk particles anymore if jamming occurs. The scaling law used to speed up the DEM simulation by reducing particle stiffness may fail for the granular flow passing through the clearance.

3) Jammed particles are almost all located in the gap between the blade tip and vessel wall. The jamming structure may only be able to resist the load from the blade in its rotational direction. The particle surface maybe damaged and worn by the blade tip during the survival time of jamming.

4) The dynamics of jammed particles are determined by the combined effects of stationary wall and blade shearing. The jamming of particles in the impeller-based powder rheometer is in low frequency with value less than 50 Hz, and the duration of single jamming event is usually less than 0.04 s.

**Acknowledgments**

The authors are grateful to the National Key Research and Development Program of China (Grant No. 2022YFB4602202) and National Natural Science Foundation of China (Grant No. 51806099). The corresponding author is thankful to Professor Mojtaba Ghadiri and Dr. Wei Pin Goh, University of Leeds, UK, for their inspirations and discussions on this work.

**Appendix**

Fig. 11 shows the variation of the blade force in the vertical direction with time in the log(y) plot, where only positive value is included. The trends in all cases are almost the same. Fig. 12 shows the probability distribution of the jammed particles. The probability decreases sharply with the increase of the number of jammed particles. Fig. 13 shows the variation of total force and torque of the blade in the vertical direction in the experiments shown in Fig. 10, in which the data is auto filtered and smoothed by the rheometer and exported at a frequency of 25 Hz.

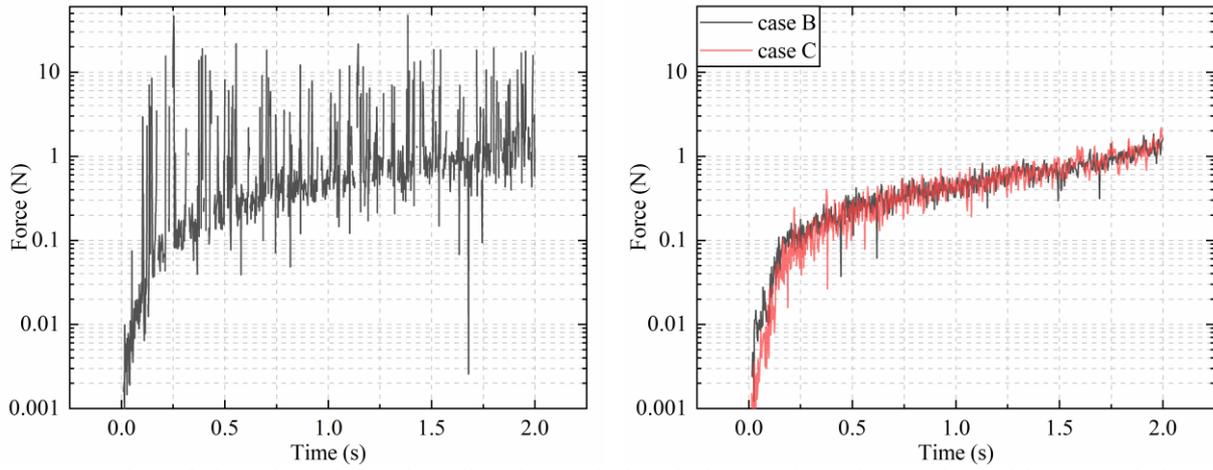

Fig. 11. Variation of blade force in vertical direction with time in log(y) plot, where only positive value is included.

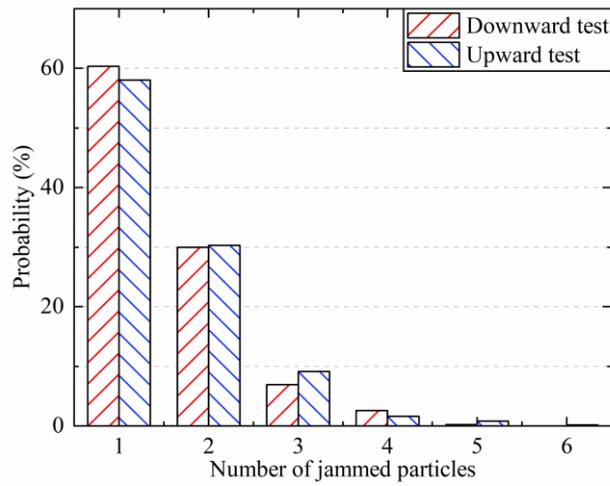

Fig. 12. Variation of the probability with the number of jammed particles

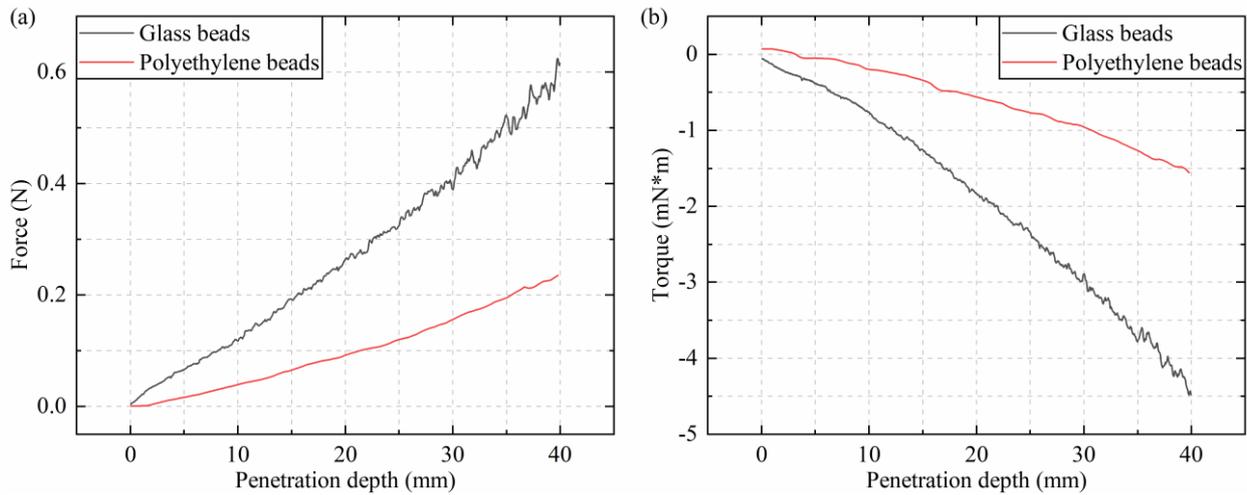

Fig. 13. Variation of total force and torque of the blade in the vertical direction in the experiments shown in Fig. 10, in which the data is auto filtered and smoothed by the rheometer and exported at a frequency of 25 Hz.